\documentclass[12pt]{iopart}
 \pdfoutput=1 
\usepackage{graphicx,color}								
\usepackage{amssymb}
\usepackage{array}
\usepackage{subfigure}
\usepackage{siunitx}

\usepackage{cite}
\begin{document}\title{Surface melting and crystallisation driven by sedimentation: a particle resolved study}

\author{Francesco Turci$^1$ and C. Patrick Royall $^1$$^,$$^2$$^,$$^3$}
\address{$^1$H.H. Wills Physics Laboratory, Tyndall Avenue, Bristol, BS8 1TL, UK}
\address{$^2$School of Chemistry, University of Bristol, Cantock's Close, Bristol, BS8 1TS, UK}
\address{$^3$Centre for Nanoscience and Quantum Information, Tyndall Avenue, Bristol, BS8 1FD, UK}

\begin{abstract}
We investigate the effects of the reversal of the gravitational field onto a sedimented and partially crystallised suspension of nearly-hard sphere colloids. We analyse the structural changes that take place during the melting of the crystalline regions and the reorganisation and assembly of the sedimenting particles. Through a comparison with numerical simulation, we access the single-particle kinetics and identify the key structural mechanism in the competition between five-fold symmetric and cubic crystalline structures. With the use of a coarse-grained, discrete model, we reproduce the kinetic network of reactions underpinning crystallisation and highlight the main microscopic transitions.
\end{abstract}

\section{Introduction}
Colloidal ``hard'' spheres are a classical model for the experimental study of the thermodynamics and the dynamics of liquids, crystals and glasses \cite{Pusey:1986dv,Palberg:2014cn,Royall:2013gs}. In particular, they constitute an excellent system for the study of crystallisation, since they allow for the identification at the particle level of the structural changes and the kinetic processes that accompany the transition \cite{Allahyarov:2015ca}, both in the bulk \cite{Ackerson:1995vy,He:1996bl,Harland:1997dr,Schilling:2010hj} and in presence of interfaces \cite{Lin:2000cc,Gasser:2001gy,Hoogenboom:2004bc,deVilleneuve:2005ih,Gasser:2009cl,Sandomirski:2011tf}. 
In principle, it is possible to prepare density matched suspensions, where gravity plays no role and crystallisation stems from homogeneous or heterogeneous nucleation, depending on the boundary conditions of the experimental sample. However, even a minor density mismatch between the fluid and the particles leads to the formation of uneven density profiles reflecting the isothermal hard-sphere equation of state \cite{Piazza:1993kg,Phan:1996gp}, and their denser regions trigger the formation of crystalline nuclei \cite{Russo:2013in}. These lead to crystallisation and the formation of time-evolving solid-liquid interfaces. Indeed, the nature of crystallisation in hard-sphere colloids is markedly different in microgravity, where dendritic growth is found \cite{Zhu:1997wc, Cheng:2001dda}. At equilibrium, in the presence of flat walls, layering and ordering in the direction of the gravity field at the surface of the walls has been observed \cite{Hoogenboom:2003jpb} and can be theoretically predicted in the framework of classical density functional theory \cite{vonGrunberg:1999gg}, while the introduction of rough, disordered walls has been shown to diminish or suppress such an effect \cite{Scheidler:2002kc,Scheidler:2004eh}. 

Although the equilibrium behaviour of sedimenting hard particles is well known \cite{Biben:1993bg,BIBEN:1994fp}, advances in the description of the dynamics at the single-particle level have been achieved only relatively recently \cite{Royall2007,Lowen:2008jw}, with a main focus on the time-evolution of the density profiles. This has revealed that a representation through the usage of a 1-dimensional density functional theory \cite{Emmerich:2012ko} approach can reproduce extremely well the experimental features, suggesting that growth takes place under quasi-equilibrium conditions. 

However, this approach cannot yet tackle situations where a fluid-solid transition occurs. Consequently, little is known about the detailed structure of the particles forming the interfaces, the role of eventual imperfections and the kinetics pathways that lead to the formation of the sedimented equilibrium profile. This aspect is important, since a better understanding of the local kinetics  would allow us to gain valuable insight on the intermediate steps through which crystallisation takes place. This would improve our understanding of interfacial growth, and improve surface properties of crystalline materials.
\hfill\break

In this work, we investigate an experimental system of nearly hard-sphere colloids confined by spontaneously formed rough walls and focus on the kinetics of sedimentation and its effects on the structure of the liquid and the crystal, under moderate density mismatch conditions. In particular, we discuss the dissolution of partly formed crystallites and the structural response of a sedimented crystal to the inversion of the gravitational field. To do so, local structures are identified and measured both in the experiments and in a compatible atomistic simulation, and a discrete, coarse-grained model for the structural differentiation is provided. We reveal that the local crystallisation mechanism is dominated by quasi-equilibrium processes, consistently with previous work \cite{Turci:2014ej,Taffs:2013kr}: this result potentially justifies the employment of techniques based on the usage of modified diffusion equations \textit{\`a la} Smoluchowski, such as dynamical density functional theory or phase field models \cite{Emmerich:2012ko,Neuhaus:2014gq, Podmaniczky:2015ge}.

This article is organised as follows: in Section 2 we describe the experimental system of nearly hard spheres and its numerical model; in Section 3 we analyse the time evolution of the experimental and simulated system, focussing first on the density profiles (Section 3.1) on the kinetics of local structure formation (Section 3.2) and the description of a coarse-grained model for the kinetic pathways leading to crystallisation or melting; we conclude in Section 4 with a critical assessment of our results.

\section{Model and  methods}
\label{sec:model}
\subsection{Crystallisation under sedimentation}
In order to control the self-assembly of colloidal  particles into crystalline states, we use gravity to manipulate the local density. Preparing a slightly density mis-matched colloidal suspension allows for the control of the driving force for crystallisation, which triggers the densification of the sample and the subsequent crystalline assembly. An inversion of the gravitational field (obtained by a vertical inversion of the prepared sample) sets the entire crystal under a considerable strain that induces a gradual melting, motion of the liquid-solid interfaces, re-assembly and settling of the crystal in a new equilibrium configuration. 

\begin{figure}[t]
	\centering
	\includegraphics[scale=1]{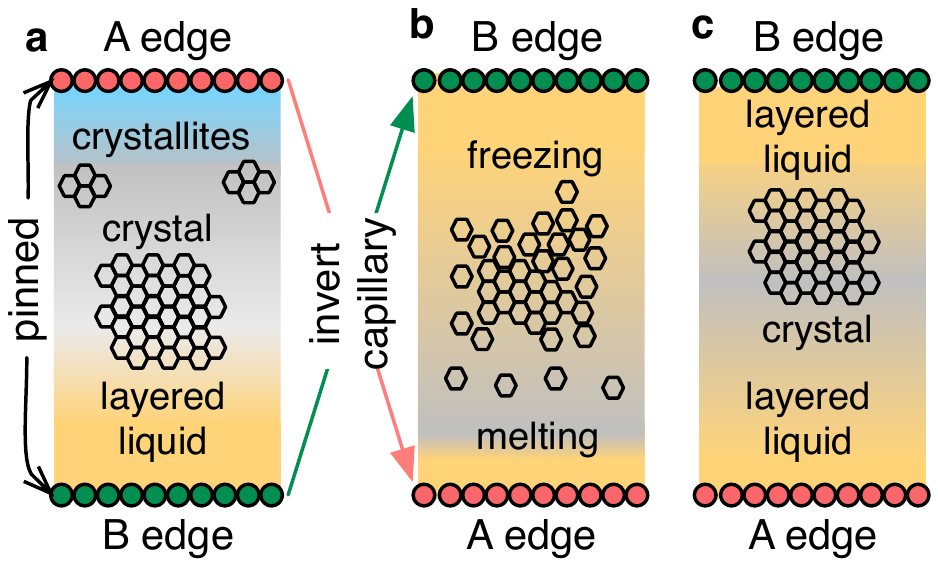}
	\caption{Schematic representation of the experimental protocol and observations. In (a), the 24-hours sedimentation has built a top layer where the crystallites are formed inside an initially layered liquid, an intermediate crystalline region and a fluid region. When reversed, (b) the system reorganises, dissolving the crystallites on the A edge, densifying  and building up order through crystallisation close to the B edge (c).}
	\label{fig:general_scheme}
\end{figure}

In our experiment, we consider a dense colloidal suspension confined in a glass capillary of rectangular section. Residual charges on the particles and substrates and Van der Waals' forces lead to the formation of disordered single layers of particles on the bottom and top surfaces of the capillary \cite{Hoogenboom:2002dn}, limiting the ordering induced by the glass confinement. It is only after the inversion of the capillary that we collect our observations and perform our analysis of the structural changes that accompany melting and re-crystallisation, see Figure \ref{fig:general_scheme}.
\subsection{Experimental and simulation details}
\begin{figure}
	\centering
	\includegraphics{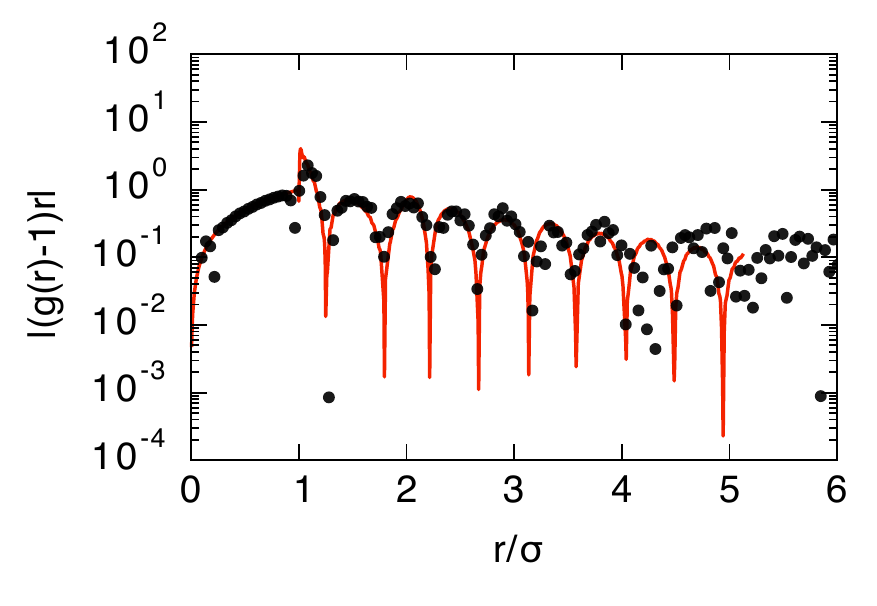}
	\caption{Determination of the particle's diameter from the decay of the experimental radial distribution function at $\phi=0.52$ compared to the corresponding expression from the Percus-Yevick approximation. A value of $\sigma=\SI{2.0}{\micro\meter}$ is found.}
	\label{fig:g_r}
\end{figure}

We employ confocal microscopy to obtain three-dimensional images of cross section $92.16\times \SI{92.16}{\micro\meter^2}=46.8\sigma\times 46.8\sigma$  of $N\approx8\cdot 10^4$ polymethyl methacrylate (PMMA) particles in a $\Delta z= \SI{100}{\micro\meter}\approx50\sigma$ thick glass capillary, where $\sigma$ is effective diameter of the particles. We image the entire height of the capillary with a LEICA SP8 confocal microscope with white light laser, at an excitation wavelength of $\SI{552}{\nano\meter}$. The dry particles have an estimated diameter of $\sigma_{\rm dry}\approx \SI{1.8} {\micro\meter}$ and polydipersity $\Delta<5\%$. When in suspension with a mixture of cyclohexyl bromide (CHB) and cis-decalin, the effective diameter shifts to $\sigma=2 a\approx\SI{2.0} {\micro\meter}$, due to electrostatic interactions that locally broaden and soften the potential (see the pair correlation function in Figure \ref{fig:g_r}), and some swelling in the solvent \cite{Poon:2012gv, Royall:2013gs}.
The experimental conditions are such that the P\'eclet number of the system is 
\begin{equation}
 	{\rm Pe}=\frac{4\pi\Delta\rho g a^3}{3 k_BT}\approx 0.31
 \end{equation}
 and the resulting gravitational length is $\xi_g=a/{\rm Pe}\approx 1.60\sigma$. The Brownian time, \textit{i.e.} the time needed to diffuse of a particle diameter $\sigma$ is approximately $\tau_B=\sigma^2/D_0=2 s$, where $D_0$ is the short-time infinite dilution diffusion coefficient.
 
 As discussed above, due to the residual charges on the particles, a single layer of particles is absorbed on the A and B edges of the capillary in the $z$ dimension. The arrangement of the pinned particles (immobile during for the whole duration of the experiment) is disordered, as illustrated in Figure \ref{fig:scheme}.a , and reduces (without suppressing in our case) the fluid layering at the edges.

In order to follow the mechanisms that lead to the phase changes during sedimentation, the experimental coordinates are also used in order to generate molecular dynamics trajectories. The dynamics is modelled with Langevin Dynamics in the over-damped limit and the interaction potential between particles of mass $m$ is chosen to model nearly-hard spheres with the combination of a hard truncated Morse potential at short radial distances and a Yukawa tail at larger distances:

\begin{equation}
	\beta u(r)=\beta\varepsilon_{M}\left[1+e^{\rho_0 (1-r/\sigma)}\left(e^{\rho_0(1-r/\sigma)
}-2\right) \right]+\beta\varepsilon_{Y}\frac{e^{\kappa(r/\sigma-1)}}{r/\sigma}.
	\label{eq:pot}
\end{equation}
We choose to truncate the Morse interaction where it vanishes ($r^{\rm cut}_{M}=\sigma$) and the Yukawa part at $r_{Y}^{\rm cut}=2\sigma$. Following \cite{Taffs:2013kr}, we adjust the parameters in Eq.\ref{eq:pot} in order to well reproduce the experimental radial distribution function of the nearly-hard sphere system: the inverse temperature $\beta=k_BT$ scales the interaction the Morse and Yukawa interaction strengths $\beta\varepsilon_{M}=1$ and $\varepsilon_{Y}=1$, the range parameter is fixed to $\rho_0=25$ while the inverse Debye screening length is set to $\kappa\sigma=30$. The resulting interaction potential combines a very hard component at distances $\sim 1\sigma$ and a residual, weak repulsion up to $2\sigma$, accounting for the imperfect charge balance at the surface of the particles.

Sedimentation is reproduced in the numerical simulation through the imposition of an external body force $F_g=m g$  where the value of $g$ accommodates with the experimental value of the gravitational length $\xi_g=k_BT/mg$.
 
Local order information from both the experimental and the simulated trajectories is obtained with the use of the Topological Cluster Classification \cite{Malins:2013jzb}: particle neighbours are found through a Voronoi tessellation and local structures are identified from a list of 33 local candidate local structures that are energy-minima for simple pair potentials such as Morse and Lennard-Jones interactions, which are therefore relevant for the present case.

\section{Crystal formation, melting and reorganisation}

\begin{figure}[t]
\centering
\includegraphics[scale=1]{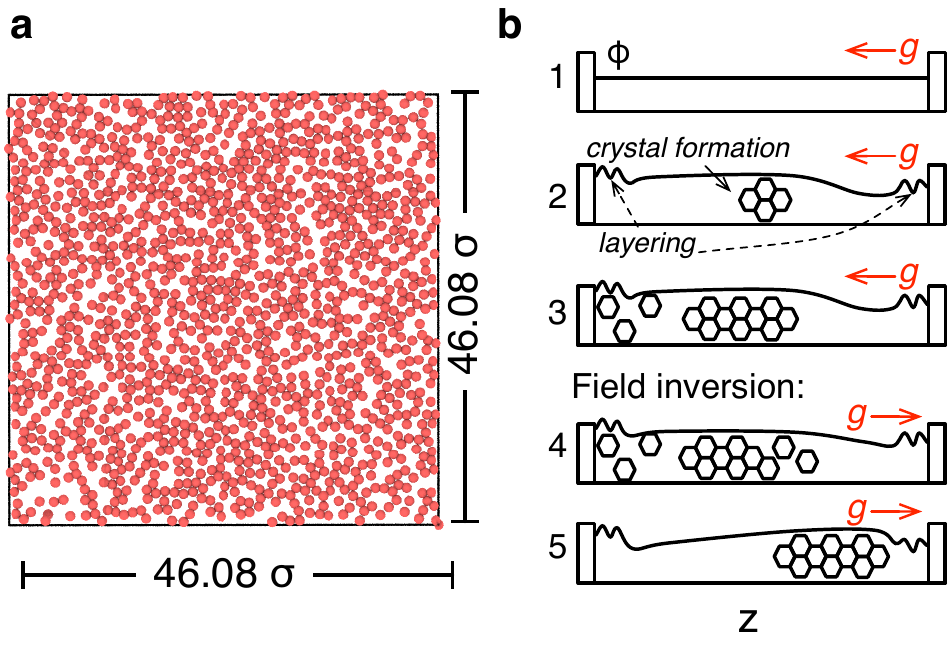}
\caption{(a) Rendering of the particles pinned on edge A of the capillary by residual charges. (b) Sketches representing the time evolution of the density profile $\phi(z)$  of the system: from an initially uniform profile (1), sedimentation begins inducing  densification and the formation of a few layers at the edges and crystallites (schematically represented as hexagons) in the dense region (2); this grows and forms a large crystal region in the large section of the capillary; when we invert the gravitational field, melting (4) is observed, followed by re-assembly into a new crystal (5).  }	
\label{fig:scheme}
\end{figure}

The experimental protocol is the following  (see Figure \ref{fig:scheme}.b): a dense suspension of PMMA particles at packing fraction $\phi=0.52(2)$ is initially dispersed and left to sediment for about 24 hours $\approx 43200\tau_B$. Subsequently the system is flipped upside-down and observed for about 13 hours $\approx 23400 \tau_B$ with confocal microscopy. This technique allows us to resolve the 3d individual particle coordinates, which are detected and analysed via a multi-scale particle-tracking algorithm \cite{Leocmach:2013cw}. The three-dimensional information allows us to identify local structures of different symmetries: we mainly focus on structures which comply with crystalline symmetries (4-folded or cubic symmetries); structures with 5-folded symmetry, incommensurate with 3d Euclidean space and therefore incapable of forming crystallites; tetrahedrally ordered structures. The state and the dynamics of the sedimenting and crystallising fluid can be assessed directly from the density profiles of the different local structures along the vertical $z$ direction (see Figures \ref{fig:expVSsim} and \ref{fig:profiles}).

From the analysis of the experimental trajectories, we draw the following observations and interpretations: On the edges of the capillary, residual surface charges fix a single layer of particles in a disordered state; Due to the moderate density mismatch between the fluid and the particles $\Delta \rho\approx\SI{0.03}{\gram/\centi\meter^3}$, a mildly layered, partly crystallised structure is formed during the 24 hours of unobserved sedimentation on side A of the capillary; this region is in contact with a bulk crystal with fewer defects; finally the crystalline region decays into an amorphous liquid, with an interface width of about $\sim5\sigma$; layering of the amorphous fluid is also observed at edge B of the capillary.
When we reverse the sample in the vertical $z$ direction and observe it for 13 hours, the gravitational force lowers the density in the originally layered region in the vicinity of the edge A and in the crystalline region, increasing it in the proximity of edge B. The interface between the original liquid region and the solid increases in density and displays further reorganisation, with a decrease of the number of defects; in the vicinity of the edge A, the remnants of the layered phase disappear and all memory of the (failed) crystalline order fades too, establishing a new liquid-solid interface of approximately $5\sigma$ in width. The system has therefore reached its new equilibrium state.

\subsection{Experimental density profiles}

\begin{figure}[t]
	\centering
	\includegraphics{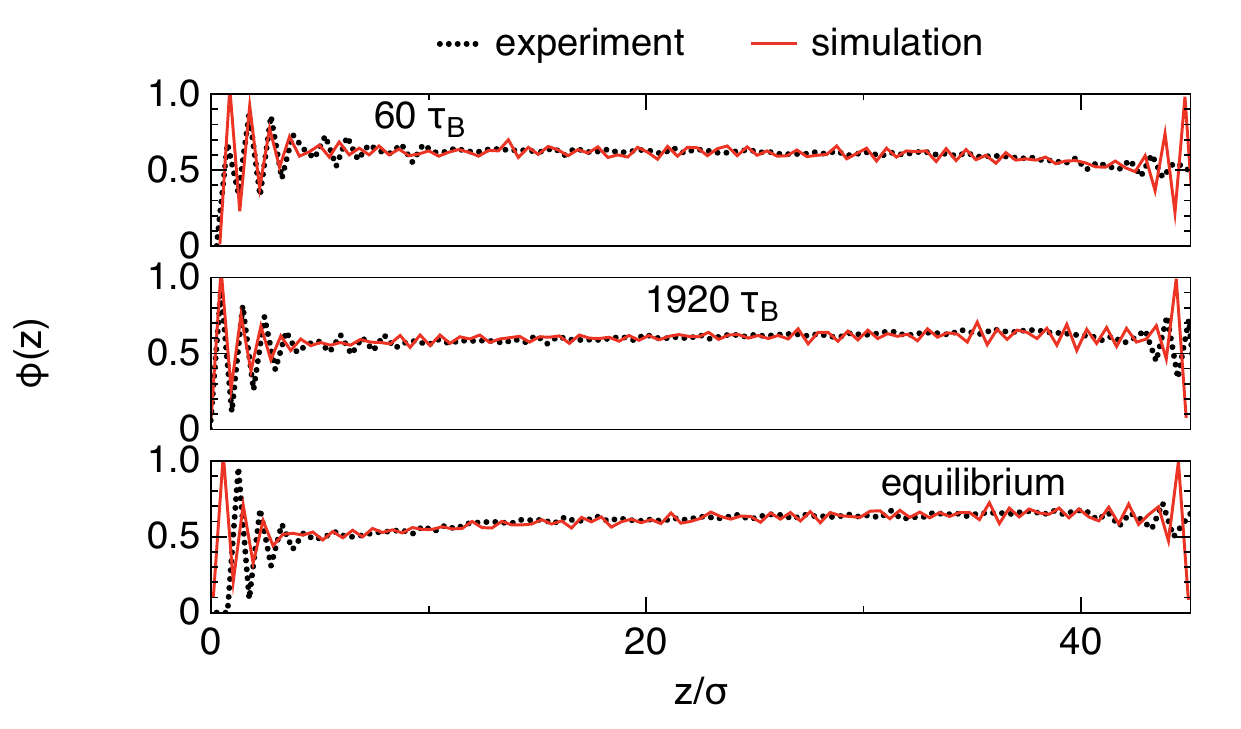}
	\caption{Packing fraction profiles from experiments and Langevin dynamics during the time evolution. Particle-tracked initial coordinates have been given in input to the simulation.}
	\label{fig:expVSsim}
\end{figure}

\begin{figure}[t]
\centering
\includegraphics[scale=1]{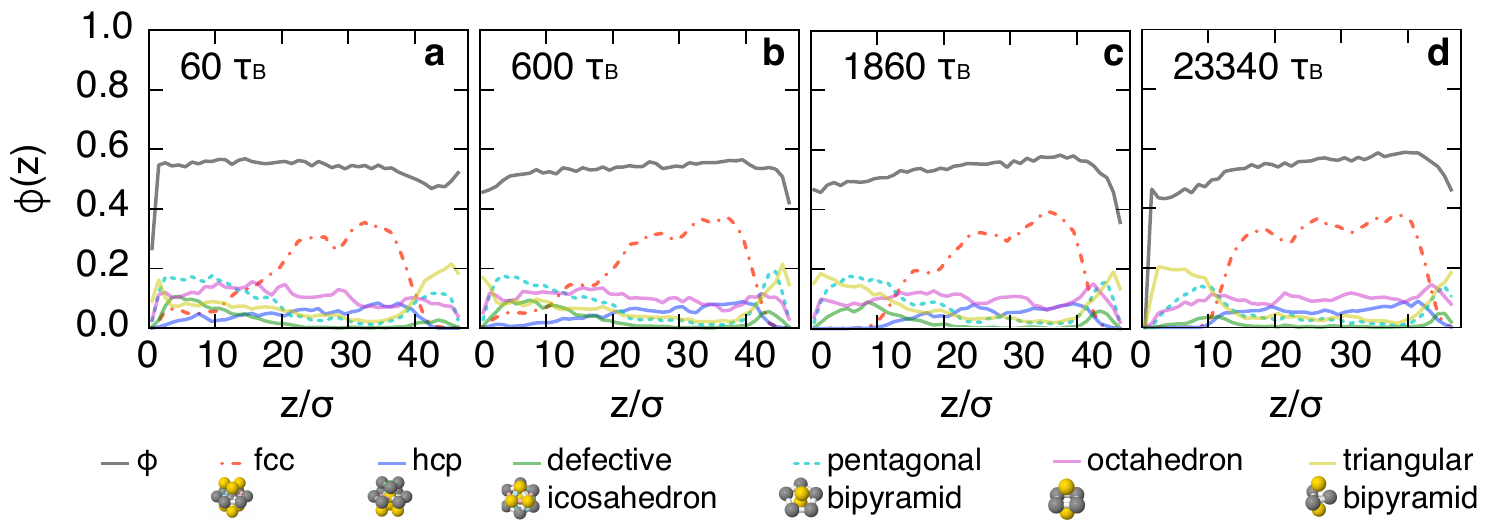}
\caption{ Experimental density and structure concentration profiles during sedimentation. At time $t=0$ the sample is reversed so that the gravitational force dissolves the crystalline state initially in contact with the $z/\sigma=0$ edge of the capillary.}	
\label{fig:profiles}
\end{figure}

From the analysis of the tracked coordinates, we follow the detailed time evolution of the system. We image the whole z-dimension of the capillary in a sequence of logarithmically spaced scans of $1,2,4,8,\SI{32}{\minute}$ after the reversal of the sample. The time evolution of the packing-fraction profiles is illustrated with dotted lines in Figure \ref{fig:expVSsim}, compared to Langevin dynamics simulations initiated form an experimentally tracked configuration (red lines). During the 13-hours observation time, the density profiles show an initial excess at the A edge of the capillary (z=0), associated to layering. The analysis in terms of local structures demonstrates that this is associated to the formation of crystallites of variable size and shape immersed in a layered liquid. In particular (Figure \ref{fig:profiles}.a) one observes an excess of five-folded symmetric structures (a 7-particle structure, the pentagonal bipyramid) in the vicinity of the A side, indicating frustration between the crystallites. A large crystal, in the face-centred cubic (fcc) phase, with many grain boundaries, is present in the middle of the capillary at more moderate densities which rapidly decay in a layered liquid, rich in small tetrahedral structures (5-particle triangular bipyramids) at the B edge of the capillary ($z\sim45\sigma$).

\begin{figure}[tb]
\centering
\includegraphics{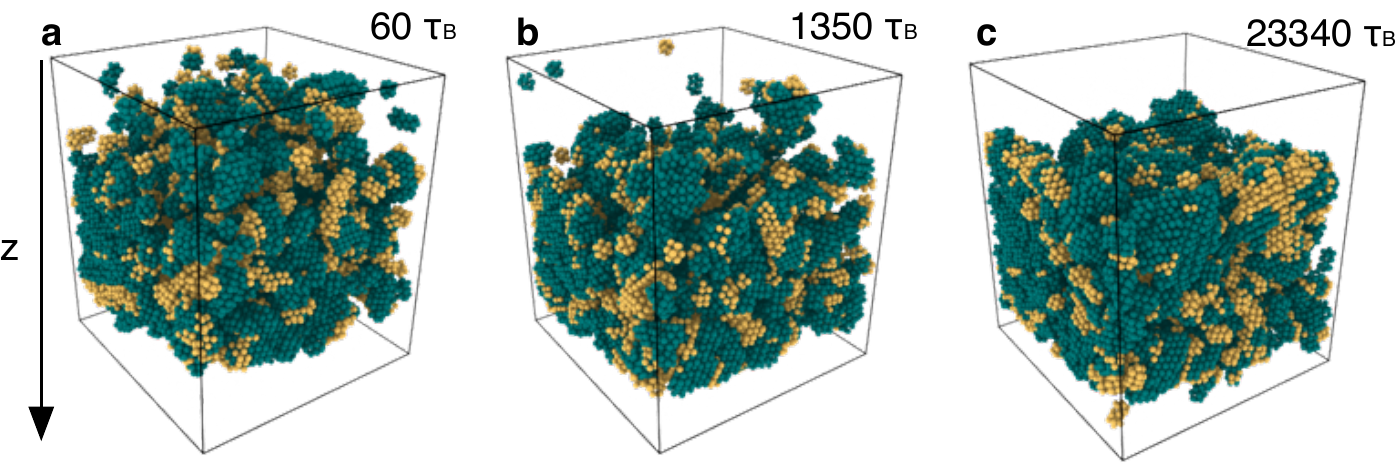}
\caption{(a-c) Experimental coordinates of particles detected in fcc  (green) and hcp (yellow) local structures during sedimentation. Particles in other arrangements are not plotted.}
\label{fig:fcchcp}	
\end{figure}

The gradual process of sedimentation leads to a reversal of the density imbalance, with a progressive increase of the packing fraction on the B edge and a reduction on the A edge, accompanied by the reorganisation of the crystal-liquid interfaces and the crystal itself. In particular, one observes (Figure \ref{fig:profiles}b-d) that the concentration of particles identified in fcc structures is enhanced in the crystalline region with the depletion of defects mainly represented by five-fold symmetric and tetrahedral structures (10-particles defective icosahedra, pentagonal and triangular bipyramids). Inspection of the 3d-coordinates (Figure \ref{fig:fcchcp}) shows that sedimentation improves the crystal quality, removing defects and grain boundaries, and forming a more compact crystal.
 
\subsection{Kinetics of the reorganisation}
\begin{figure}[t]
\centering
\includegraphics{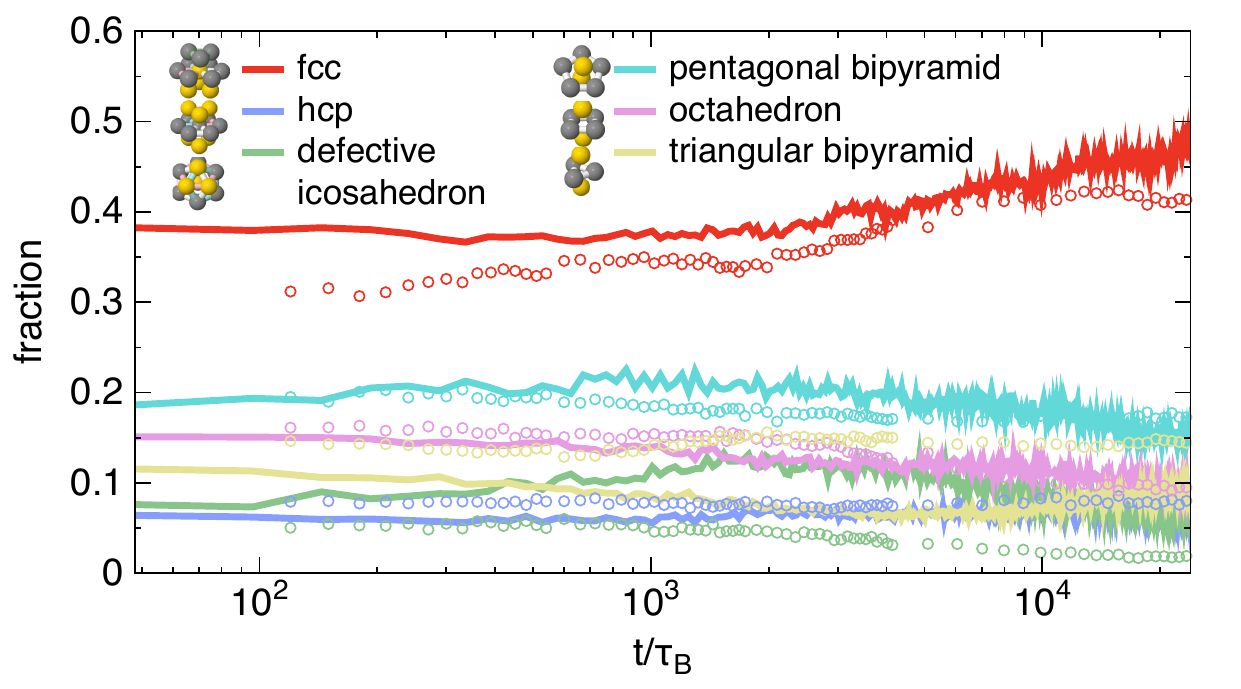}
\caption{Time evolution of the structure concentrations: experiment (circles) and simulations (continuous lines). Qualitative and quantitative agreement is observed; the concentration of the defective icosahedra shows the biggest quantitative differences.}
\label{fig:timev}	
\end{figure}

The limited time resolution of the experiments means that we can image the sample at most every $\approx 30\tau_B$ and it hinders a thorough analysis of the kinetic pathways  leading to the melting of the defective crystal and the crystallisation of the liquid region. In particular we cannot track the individual particles and fully reconstruct their kinetic pathways. In order to overcome these limitations, a Langevin Dynamics simulation is restarted from the initial coordinates tracked from the experiment, with carefully chosen parameters matching the sedimentation conditions of the system (as previously described in Section \ref{sec:model}). This allows us to follow the trajectory of all the particles during the sedimentation process, and identify that local structural changes that every region of the system undergoes.
As demonstrated in Figure \ref{fig:expVSsim} the numerical simulation closely reproduces the outcomes of the experiment, following the time evolution of the experimental sedimentation profiles and attaining a comparable equilibrium state. Furthermore, Figure \ref{fig:timev} demonstrates that the time evolution of the local structure in the experiment is \textit{qualitatively} and \textit{quantitatively} captured by the Langevin Dynamics, with the biggest discrepancies being associated to the five-fold symmetric local structures. This suggests that long-range hydrodynamic interactions, intrinsically present in the experimental system but systematically neglected in the numerical model, may play only a minor role in the reorganisation of the fluid and the crystal, eventually reducing the likelihood of formation of the most complex 5-fold symmetric structures (\textit{i.e.} the defective icosahedra). While at lower densities non-uniform colloidal suspensions have been shown to be characterised by hydrodynamic instabilities \cite{Wysocki:2010kc,Wysocki:2009ky}, in our much denser system the crystal growth (and melting) appears to be dominated by local, short-range interactions. 

\subsection{Coarse-grained model}
We use the statistics from simulated trajectories in order to identify the structural mechanisms that lead to the phase changes at the liquid-solid interfaces. To do so, we employ the structural order parameters provided by the Topological Cluster Classification to build a coarse grained model describing the thermodynamic state of the sedimenting system, with seven discrete states associated to three main families: the cubic order (hcp, fcc structures and the octahedra, based on four-membered rings), the five-fold symmetric structures (pentagonal bipyramid and defective icosahedra) that play a major role in frustrated hard spheres, small triangular bipyramids (bicapped tetrahedra) or liquid structures smaller than tetrahedra (termed \textit{other}). This is conceptually analogous to coarse-grained approaches used to investigate crystallisation for sheared colloidal suspensions \cite{Lander:2013jk}, with a finer distinction between the pre-crystalline states. 

As suggested by the density profiles, tetrahedra and five-fold symmetric structures are well represented in the liquid region while the cubic order is characteristic of the solid region.

\begin{figure}[t!]
	\centering
	\includegraphics{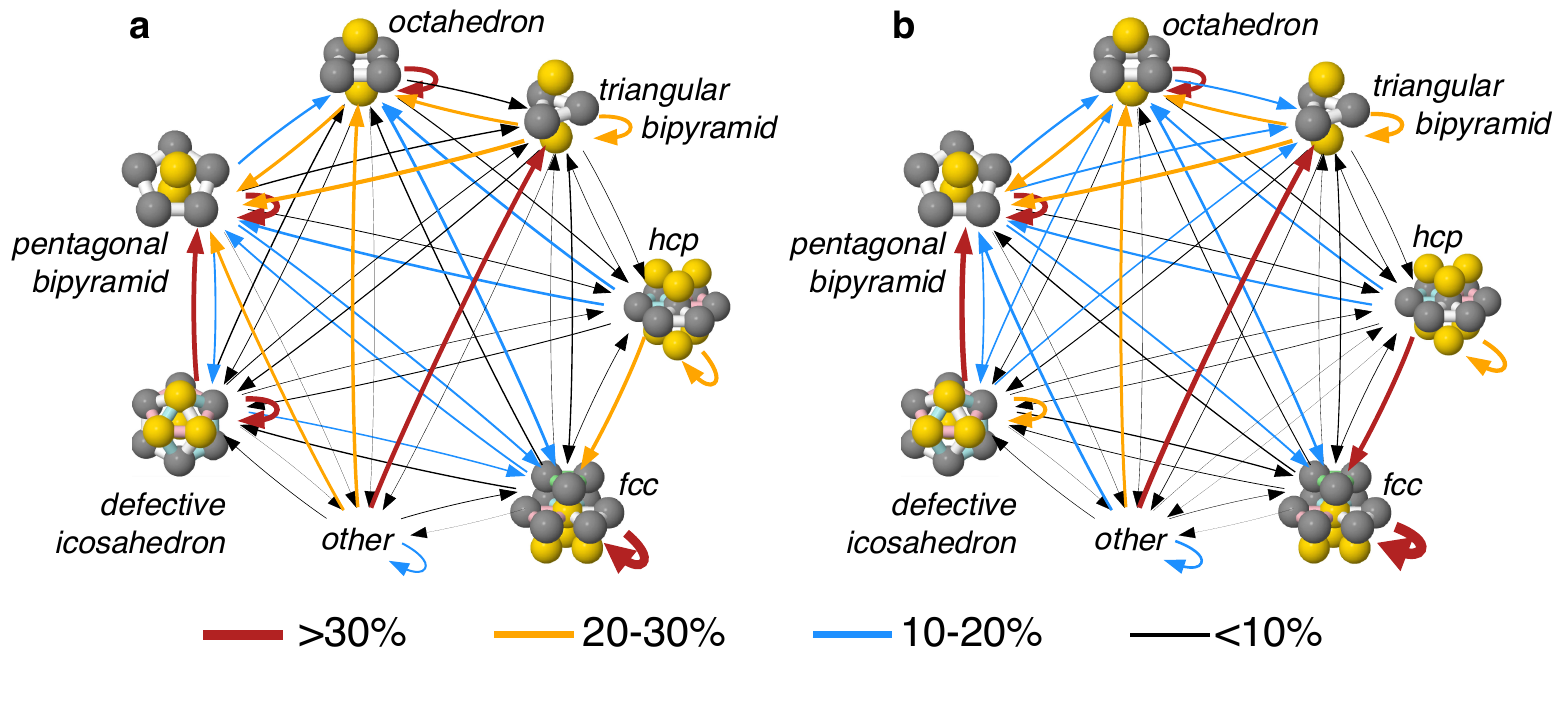}
	\caption{Directed graphs representing the transition matrices (a) $T^{\rm noneq}$ for the particles changing local structure during the sedimentation transient and (b) $T^{\rm eq}$ for the steady state regime. The thickness and the colours of the edges are proportional to the transition probabilities, as indicated in the legend. The local structures are represented with differently coloured particles in order to highlight the ring structures (triangular, square and pentagonal rings). }
	\label{fig:graphs}
\end{figure}

From the Langevin Dynamics simulations we extract the individual particle trajectories, and associate at any time a discrete state to the particles, chosen from the set of selected structures. Then, we count the transitions from one state to another and estimate transition probabilities $p_{ij}$ from state $i$ to state $j$. We focus on two transition matrices derived from different estimates of the probabilities: i) the matrix $T^{\rm eq}$ where the probabilities are estimated from \textit{all} the particles transitions after time $t_{\rm eq}$, once the stationary equilibrium profile is attained; ii) the matrix $T^{noneq}$ where the probabilities are estimated only during the transient before equilibrium and for particles that reside in a crystalline phase for less than half of the equilibration time $t_{\rm eq}$. These matrices are represented in Figure \ref{fig:graphs} as directed graphs where the thickness and the colour of the lines correspond to the magnitude of the transition probabilities.

The graphs lead to the following observations: In the transient regime [Figure \ref{fig:graphs}.(a)], local structures of less than 5 particles (named "others") mainly transform into triangular bipyramids (tetrahedra); such local environment can either take the five-fold symmetric route of the pentagonal bipyramids and defective icosahedra (where a feedback mechanisms makes the bigger 10-particle defective icosahedra mainly decay in the smaller 7-particle pentagonal structure) or the cubic route, through a conversion into octahedra acting as precursors that finally form fcc regions. In the steady state matrix computed for all the particles, Figure \ref{fig:graphs}.b, such a tetrahedral route is confirmed and the overall structure of the network is only mildly modified, suggesting that under the moderate density imbalance, the growth mechanism is a quasi-equilibrium one \cite{Turci:2014ej}. We notice that the role of the hexagonal close packed ordering (hcp) is marginal in both the two regimes and that it does not appear to represent a main route to the formation of fcc structures.

\begin{figure}[t]
	\centering
	\includegraphics{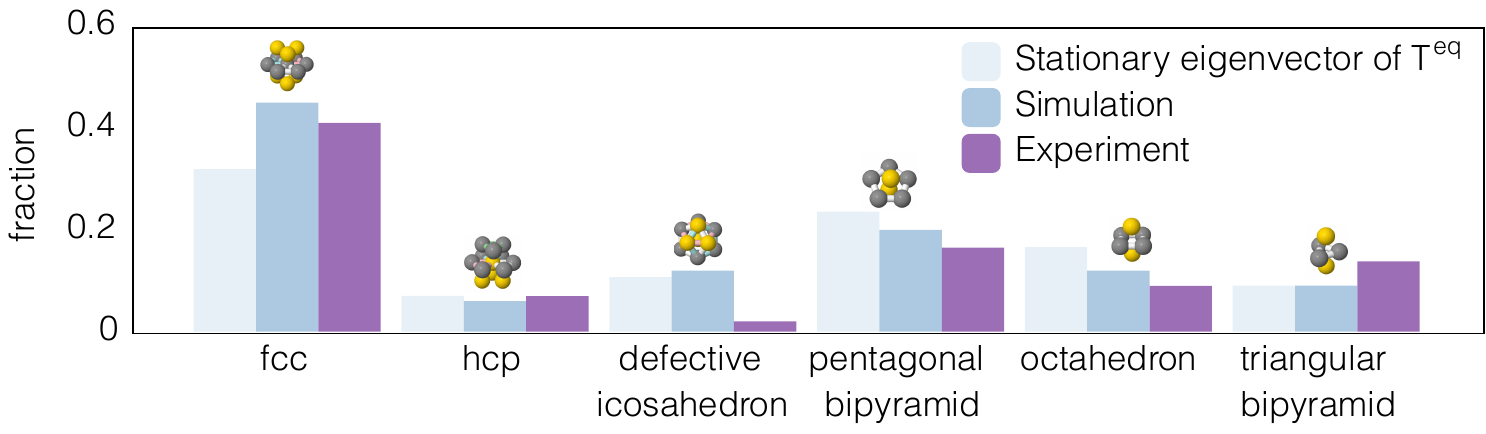}
	\caption{Comparison between the steady state concentrations of different local structures according to the the transition matrix $T^{\rm eq}$, the numerical simulations and the experimental results.}
	\label{fig:perc}
\end{figure}

Diagonalising the steady state state matrix $T^{\rm eq}$ and identifying the eigenvector associated to the largest eigenvalue, one obtains estimates for the expected steady state concentrations for the discrete model represented in the graphs. Notice that this approach involves a radical reduction of the complexity of the problem, since no spatial information is encoded in the graph itself: yet, the steady state values correctly reproduce the hierarchy of the final concentrations, with maximum difference of $\sim10\%$ in concentration. The steady-sate probabilities resulting from the experiment, the simulation and the coarse-grained model are illustrated in Figure \ref{fig:perc}, showing that the biggest discrepancy occurs in the estimated concentration of the ten-particle defective icosahedron, essentially suppressed in the experiment.
 
\section{Conclusions}
We have performed experiments and simulations on the sedimentation of colloidal nearly-hard spheres. Melting and re-assembly have been observed in experiment and simulations and described in terms of local structural signatures and their time evolution. We have highlighted the competition between crystalline cubic order and the five-folded symmetry, illustrating that crystallisation progresses with the gradual suppression of the latter and the formation of small precursors. This is consistent with previous observations of five-fold symmetric order parallel to the solid-liquid interface of dense liquids in contact with structured substrates \cite{Reichert:2000kl,Heni:2002bc}. Employing Langevin dynamics numerical simulations, we managed to reproduce the overall kinetic and structural changes occurring in the experiment, suggesting that the contribution of long range hydrodynamic interactions to the determination of the surface melting and crystallisation mechanisms may be limited.
Finally, the kinetic pathways reproduced in the numerical simulation allowed us to propose a simplified coarse-grained model of the structural transitions, that sheds light on a possible hierarchy for the melting and crystallisation kinetics. This approach may be extended in future in order to improve spatially coarse-grained models and predict local structural changes.

\section*{Acknowledgements}

FT thanks Julien Lam and Nicholas Wood for their precious advice. CPR acknowledges the Royal Society for funding and Kyoto University SPIRITS fund. FT and CPR acknowledge the European Research Council (ERC consolidator grant NANOPRS, project number 617266).

\vspace{2cm} 

\bibliographystyle{iopart-num}
\providecommand{\newblock}{}


\end{document}